\documentclass[onecolumn,showpacs,preprintnumbers,floatfix,amsmath,amssymb,aps]{revtex4}
\usepackage{bm,tikz}
\usepackage{color}

\usepackage{graphicx,subfig,epsfig,epsf,color}
\usepackage[utf8x]{inputenc}
\usepackage{amssymb,amsmath,appendix}
\usepackage{slashed}
\usepackage{array}
\usepackage{hyperref}
\usepackage{ulem}

\begin{document}
\title{Impact of dark boson mediated feeble interaction between dark matter and hadronic matter on $f$-mode oscillation of neutron stars}

\author{Debashree Sen$^{1}$ and Atanu Guha$^2$}

\affiliation{$^1$Center for Extreme Nuclear Matters (CENuM), Korea University, Seoul 02841, Korea}
\affiliation{$^2$Department of Physics, Chungnam National University,\\ 99, Daehak-ro, Yuseong-gu, Daejeon-34134, South Korea}

\email{debashreesen88@gmail.com, atanu@cnu.ac.kr}

\date{\today}




\begin{abstract}

We studied the possible presence of dark matter (DM) in neutron stars (NSs) and the structural properties of the DM admixed NSs (DMANSs) in one of our recent works \cite{Guha:2024pnn}. The feeble interaction between the fermionic DM ($\chi$) with the hadronic matter is introduced through a dark scalar ($\phi$) and a dark vector ($\xi$) boson as mediators. The allowed range of the mass of the fermionic DM ($m_{\chi}$), for a particular range of DM Fermi momentum ($k_F^{\chi}$), was obtained in the same work \cite{Guha:2024pnn} with respect to the various astrophysical constraints on the structural properties of compact stars viz. the mass, radius and tidal deformability. The present work is dedicated to the calculation and study of non-radial oscillation of the DMANSs using Cowling approximation. We particularly investigate the effect of presence of DM on the fundamental ($f$) mode oscillation frequencies of the DMANSs utilizing the previously obtained range of $m_{\chi}$ for four different hadronic models. In this work we thoroughly investigate how the individual and combined effects of $m_{\chi}$ and $k_F^{\chi}$ affect the $f$-mode oscillation frequency. Within the framework of our DMANS models, for a particular value of $k_F^{\chi}$, the range of $f_{max}^{DMANS}$ with respect to the allowed range of $m_{\chi}$, is also obtained in the present work for four different hadronic models. Since in the present era, the 1.4 and 2.01 $M_{\odot}$ NSs are of special interest after the detection of GW170817 and PSR J0740+6620, we particularly investigate, for the four hadronic models, the range of $f_{1.4}^{DMANS}$ and $f_{2.01}^{DMANS}$ with respect to the acceptable range of $m_{\chi}$ corresponding to the range of $k_F^{\chi}$. 
\end{abstract}




\maketitle



\section{Introduction}
\label{Intro}

Of the total energy budget of the Universe, the lion's share is made up of dark energy (about $70\%$) while the succeeding leading component is dark matter (around $25\%$). The so-called luminous matter or the baryonic matter makes up only $5\%$ of the total energy content. At present, this is an unambiguously evident fact supported by certain astrophysical and cosmological observations~\cite{Planck:2015fie,Bertone:2004pz,Aghanim:2018eyx,Bauer:2020zsj}. Till date several dedicated search methods have been implemented to obtain an idea about the interaction strength between the dark matter (DM) and standard model (SM) particles. The popular experimental search avenues, that include the direct and the indirection detection strategies~\cite{Lisanti:2016jxe,Profumo:2013yn}, have shown significant developments. The most stringent constraints obtained till date are from the leading direct detection experiments like SuperCDMS~\cite{SuperCDMS:2018mne}, XENONnT~\cite{XENON:2022ltv}, PandaX-II~\cite{PandaX-II:2020oim}, DarkSide-50~\cite{DarkSide:2018ppu}, SENSEI~\cite{Crisler:2018gci} and LUX-ZEPLIN~\cite{LZ:2022lsv}. Experiments dedicated to indirect search are FERMI-LAT~\cite{Fermi-LAT:2009ihh}, IceCube~\cite{IceCube:2014stg}, PAMELA~\cite{PAMELA:2008gwm, PAMELA:2013vxg}, AMS-02~\cite{AMS:2014bun}, Voyager~\cite{Boudaud:2016mos} and CALET~\cite{CALET:2017uxd,Adriani:2018ktz}.

Significant advancement in the phenomenological aspects has also been made parallel to the progress of the experimental probes. One of the well motivated methods to understand the properties of the DM particles and its interaction with baryonic matter is to explore the possible presence of DM in neutron stars (NSs) because NSs are highly gravitating objects that are capable of accreting matter from its surroundings including DM. Inside the NS core, matter is preserved at extreme conditions of density, pressure, compactness and gravity. Fortunately, they are accessible by direct observations. Therefore NSs serve the purpose of being unique natural astrophysical laboratories where we can investigate theoretically the properties of matter under extreme conditions. Several mechanisms can be responsible for the possible presence of DM in NSs, e.g., capture or accretion of DM halo particles by NSs~\cite{Baryakhtar:2017dbj, Joglekar:2019vzy, Bell:2019pyc, Joglekar:2020liw, Bramante:2023djs}, creation of its own DM through dark decays of neutrons~\cite{Husain:2022bxl, Husain:2022brl, Husain:2023fwb, Zhou:2023ndi, Shirke:2024ymc} or inheritance of DM from the supernovae~\cite{Nelson:2018xtr} etc. The DM particles end up being gravitationally bound to the star and the DM particles attain thermal equilibrium among themselves due to the self interactions \cite{Bell:2019pyc, Bell:2020lmm}. This amply justifies the consideration of the DM particle density $\rho_{\chi}$ to be almost constant throughout the radius of the star \cite{Panotopoulos:2017idn,Guha:2021njn,Sen:2021wev}. The DM particles thus become structural part of the NSs. This allows us to explore the effect of presence of DM on the structural properties of NSs \cite{Shakeri:2022dwg, Karkevandi:2021ygv, Lenzi:2022ypb,Lourenco:2022fmf,Sen:2021wev, Guha:2021njn, Guha:2024pnn}. 

Within NS cores, DM may not interact with hadronic matter and the two types of matter coexist in the two fluid form \cite{Lopes:2018oao, Ellis:2018bkr, Li:2012ii, Tolos:2015qra, Deliyergiyev:2019vti, Mukhopadhyay:2016dsg, Jimenez:2021nmr, Panotopoulos:2018ipq, Leung:2022wcf, Karkevandi:2021ygv, Lourenco:2021dvh, Gleason:2022eeg, Dengler:2021qcq, Panotopoulos:2018joc, Miao:2022rqj, Rutherford:2022xeb, Barbat:2024yvi, Karkevandi:2024vov, Zhen:2024xjc} whereas the interaction between the DM and the baryonic matter is also suggested by \cite{Panotopoulos:2017idn, Bertoni:2013bsa, Nelson:2018xtr, Bhat:2019tnz, Lourenco:2022fmf, Dutra:2022mxl, Hong:2023udv, Mu2023}, mostly via the Higgs boson as mediator. In case the DM interacts with hadronic matter, the interaction must be extremely weak \cite{Zheng:2016ygg} to prevent the collapse of the star into a black hole. Therefore we invoked feeble interaction between hadronic and fermionic DM $\chi$ via a new scalar mediator $\phi$ in \cite{Sen:2021wev} and also a dark vector mediator $\xi$ in \cite{Guha:2021njn, Guha:2024pnn} in order to explain the possible existence of forming DM admixed NSs (DMANSs). $\phi$ and $\xi$ interact with the hadronic matter $\psi$ with a very feeble coupling strength. The masses of DM fermion ($m_{\chi}$) and the mediators ($m_{\phi}$ and $m_{\xi}$) and their couplings ($y_{\phi}$ and $y_{\xi}$) are consistent with the self-interaction constraint from the Bullet cluster observations and from the present day relic abundance of DM, respectively. We concluded that mass of DM plays a very important role in determining the structural properties of DMANSs. The massive the DM, the less are the maximum mass, radius and tidal deformability of the DMANSs. In \cite{Guha:2024pnn} within a certain range of the constant DM particle density $\rho_{\chi}$, we determined a possible range of $m_{\chi}$ for which the DMANSs satisfy the constraints on the structural properties of compact stars obtained from PSR J0740+6620 \cite{Fonseca:2021wxt, Miller:2021qha, Riley:2021pdl}, gravitational wave (GW170817) data \cite{LIGOScientific:2018cki} and the NICER data for PSR J0030+0451 \cite{Riley:2019yda, Miller:2019cac}. For the purpose, we considered certain relativistic mean field (RMF) models to account for the hadronic matter.

In \cite{Guha:2024pnn} the suitable range of $m_{\chi}$ consistent with the astrophysical constraints, was determined by calculating the structural properties like the mass, radius and tidal deformability of the DMANSs. In the present work we focus on the phenomenon of non-radial NS oscillation and particularly on the fundamental or $f$-mode frequency of the oscillation of the DMANSs. One of the important reasons why the non-radial pulsations of NSs are of current interest is because they are connected with gravitational radiation. The oscillation of NSs is accompanied by emission of gravitational waves (GWs) of different frequencies viz., fundamental ($f$), pressure ($p$), rotational ($r$), space-time ($w$) and gravity ($g$) modes \cite{Kokkotas:1999bd}. These modes are classified on the basis of the type of the restoring force that bring the star to equilibrium. For example, in case of the $f$ and $p$ modes, the equilibrium is restored by fluid pressure while for the $g$-mode gravity restores the equilibrium. The the oscillation spectra can be an useful probe to understand the interior and properties of NSs. This is often termed as GW asteroseismology. Among the different frequencies, the most prominent $f$-mode is of special interest because the future GW detectors such as the Einstein Telescope, the Cosmic Explorer, and the LIGO O4 run, are being specially designed to match the value of $f$-mode frequency ($1-3$ kHz). Therefore, it may be expected that the $f$-mode frequency of GWs will be detected by these upcoming GW detectors. This will improve our understanding regarding the interior of NS and its composition and also for further constraining the EoS of NSs. Moreover, universal relations between the $f$-mode and compactness \cite{Andersson:1997rn}, moment of inertia \cite{Lau:2009bu} and static tidal polarizability \cite{Chan:2014kua,Sotani:2021kiw} for NSs, quark stars and hybrid stars \cite{Zhao:2022tcw} have emerged due to well corelation between the $f$-mode and these properties of compact stars. However, $p$-modes show very weak correlation with such NS properties \cite{Andersson:1997rn}. Also, the $f$-mode values are not much affected by the crust \cite{VasquezFlores:2017tkp, Pradhan:2020amo} while the $p$-mode is sensitive to the low-density crustal EoS \cite{Kunjipurayil:2022zah}. The estimation of the gravitational-radiation flux from NSs thus requires the theory of non-radial oscillations. The non-radial oscillation of NS was first developed in the framework of general relativity in \cite{1967ApJ...149..591T} while the first integrated numerical solution of the NS oscillation was derived in \cite{Lindblom:1983ps}. Later the method was simplified with the help of Cowling approximations \cite{Cowling:1941nqk} by neglecting the metric perturbations in \cite{Sotani:2010mx}. The calculation the oscillation frequencies in full GR conditions that include the spacetime oscillation also give us the opportunity to calculate the $w$-mode frequency along with the $f$ and $p$ mode frequencies. However, the gross qualitative results remain unaffected although the magnitude of frequencies in the Cowling approximation differ from those in the total GR calculations (without Cowling approximation) by up to 30\% for the $f$-mode, and $\sim$15\% for $p_1$-mode \cite{Kunjipurayil:2022zah}. A lot of studies have been done on these different oscillation modes \cite{Flores:2013yqa,Andersson:2019mxp,Zhao:2022tcw,Kumar:2024jky,Ranea-Sandoval:2018bgu,VasquezFlores:2019eht,Pradhan:2022vdf,Jyothilakshmi:2024zqn,Das:2021dru, Hong:2023udv}. Only a few of them \cite{Das:2021dru, VasquezFlores:2019eht, Hong:2023udv, Liu:2024rix, Shirke:2024ymc} particularly focused on the non-radial $f$-mode oscillations of DMANSs and \cite{Liu:2024rix} used the bound on tidal deformability from GW data in order to extract the DM composition of the star for the extracted range of $f$-mode values. However, interaction between DM and NS matter is not considered in \cite{Liu:2024rix} and therefore unlike the present work they have considered the two-fluid approach to obtain the structural properties of the DMANSs.

In the present work we intend to show how the presence of DM and its interaction with hadronic matter affects the most prominent $f$-mode frequency of the oscillation of the NSs. For the purpose we consider four well-known RMF models viz. NL3 \cite{Lalazissis:1996rd}, GM1 \cite{Glendenning:1991es}, DD2 \cite{Typel:2009sy}, and DD-ME2 \cite{Lalazissis:2005de} to describe the hadronic matter content of the star. Within a range of constant DM Fermi momentum $k_F^{\chi}$, we have already obtained the suitable range of $m_{\chi}$ required to obtain reasonable DMANSs configurations with these four hadronic models \cite{Guha:2024pnn}. Therefore we use the same in the present work in order to study the effects of presence of DM on the $f$-mode oscillation of the DMANSs in terms of $m_{\chi}$ corresponding to constant $k_F^{\chi}$ or constant $\rho_{\chi}$.

We organize the present work as follows. In the next Sec.~\ref{Formalism}, we briefly address the mechanism of invoking feeble interaction of DM with hadronic matter via the dark mediators $\phi$ and $\xi$ using the four RMF hadronic models NL3, GM1, DD2, and DD-ME2. We also discuss the methodology of estimating the structural properties like mass, radius, tidal deformability and the $f$-mode oscillation of the DMANSs. We then present the results of our estimations based on Sec.~\ref{Formalism} and corresponding discussions in Sec.~\ref{Results}. We summarize and conclude in the final section \ref{Conclusion} of the paper.


\section{Formalism}
\label{Formalism}

\subsection{Dark matter admixed neutron star models}
\label{Models}

For the $\beta$-equilibrated hadronic NS matter we consider four well-known RMF models. Of them, NL3 \cite{Lalazissis:1996rd} and GM1 \cite{Glendenning:1991es} are with non-linear self couplings while DD2 \cite{Typel:2009sy} and DD-ME2 \cite{Lalazissis:2005de} have density-dependent couplings following the Typel-Wolter ansatz \cite{Lu:2011wy}. The values of the couplings and the saturation properties of all the four hadronic models considered in this present work, can be found in the respective references and also in \cite{Xia:2022dvw}. In \cite{Guha:2024pnn} we already invoked feeble DM
interaction with hadronic matter using these four RMF models. A phenomenological treatment was considered to describe the self-interaction of non-relativistic DM by a Yukawa potential \cite{Tulin:2013teo}.  The interaction of the dark fermion ($\chi$) with the hadronic matter ($\psi$=n, p) is mediated by the scalar ($\phi$) and vector ($\xi$) new physics mediators. $\phi$ and $\xi$ interact with the hadronic matter $\psi$ with a very feeble coupling strength $g_{\phi}=g_{\xi}\sim$10$^{-4}$. Also, $\phi$ and $\xi$ have their respective couplings  with $\chi$ as $y_{\phi}$ and $y_{\xi}$.   As considered in our previous works \cite{Sen:2021wev, Guha:2021njn, Sen:2022pfr}, the values of $m_{\chi}$, $m_{\phi}$ and $m_{\xi}$ are consistent with the self-interaction constraints from bullet cluster \cite{Randall:2007ph, Bradac:2006er, Tulin:2013teo, Tulin:2017ara, Hambye:2019tjt} while the self-interaction couplings are also chosen by reproducing the observed non-baryonic relic density \cite{Belanger:2013oya, Gondolo:1990dk, Guha:2018mli}. The values of $m_{\phi}$ and $m_{\xi}$ corresponding to the range of $m_{\chi}$ are already shown in our previous works \cite{Guha:2021njn, Sen:2021wev, Sen:2022pfr}. The DM number density $\rho_{\chi}$ is considered to be constant via constant DM Fermi momentum $k_F^{\chi}$ throughout the radial profile of the star following \cite{Panotopoulos:2017idn, Guha:2024pnn, Guha:2021njn, Sen:2021wev}. It is assumed that the DM density is $10^{-3}$ times the baryon density \cite{Panotopoulos:2017idn}. The detailed methodology for obtaining the equation of state (EoS) i.e, the energy density $\varepsilon$ and pressure $P$ (as functions of baryon density $\rho$) of DMANSs using these four models can be found in \cite{Guha:2024pnn}. 

It is well-known that the crust of the NS plays important role in determining the overall radius of the NS. In our earlier works \cite{Guha:2024pnn,Guha:2021njn,Sen:2021wev} and thus in the present work we have considered constant number density of DM along the radius profile of the star following \cite{Panotopoulos:2017idn}. This consideration also implies that DM should also be present in the crust although several works on DMANSs have not considered the presence of DM in the crust \cite{Das:2018frc, Abac:2021txj}. The methodology adopted for invoking DM interaction with the NS matter in our earlier \cite{Guha:2024pnn,Guha:2021njn,Sen:2021wev} and present works, requires a Lagrangian formulation of the DMANS matter. For consistency between core and crust, we require a Lagrangian formulation of the crust matter of NS in order to obtain DM admixed crust EoS including the interaction between DM and NS crust matter through dark mediators. However, this  Lagrangian formulation of the crust matter of NS is not very common in the present literature. Therefore the calculation of the DM admixed EoS for the crust is somewhat not easy and beyond the scope of the present work. Therefore we consider DM interaction only in the core and obtain the DMANS EoS for the core.

\subsection{Structural Properties of Dark matter admixed neutron stars}

Using the DMANS EoS, we estimate the structural properties like the gravitational mass ($M$) and the radius ($R$) of the DMANSs. The metric for spherically symmetric star in static conditions is given as

\begin{eqnarray}
ds^2=-e^{2\Phi(r)}dt^2 + e^{2\Lambda(r)}dr^2 + r^2d\theta^2 + r^2 sin^2\theta d\phi^2
\label{metric}
\end{eqnarray}

Based on this metric the Tolman-Oppenheimer-Volkoff (TOV) equations are derived \cite{Tolman:1939jz,Oppenheimer:1939ne} as 

\begin{eqnarray}
\frac{dP(r)}{dr}=-\Big(\varepsilon(r)+P(r)\Big)\frac{d\Phi(r)}{dr}
\label{tov}
\end{eqnarray}

\begin{eqnarray}
\frac{d\Phi(r)}{dr}=\frac{M(r)+4\pi r^3 P(r)}{r\Big(r-2 M(r)\Big)}
\label{tov2}
\end{eqnarray}

\begin{eqnarray}
\frac{dM(r)}{dr}= 4\pi r^2 \varepsilon(r),
\label{tov3}
\end{eqnarray} 

where, $\Phi(r)$ and $\Lambda(r)$ are metric functions with respect to $r$. The mass function $M(r)=r(1-e^{-2\Lambda(r)})/2$ satisfies Eq. (\ref{tov3}).

These Eqs. (\ref{tov})-(\ref{tov3}) signify the hydrostatic equilibrium between gravity and the internal pressure of the star.

The methodology to calculate the non-radial oscillations of the NSs using the Cowling approximations, is well depicted in \cite{Sotani:2010mx}. Once we obtain the mass and radius of the star by solving the TOV Eqs. (\ref{tov}) - (\ref{tov3}), the two following coupled differential equations are solved to obtain the oscillation mode frequencies :

\begin{eqnarray}
\frac{dW(r)}{dr}=\frac{d\varepsilon(r)}{dP(r)}\Bigg[\omega^2r^2e^{\Lambda(r)-2\Phi(r)}V(r) + \frac{d\Phi(r)}{dr}W(r)\Bigg] - l(l+1)e^{\Lambda(r)}V(r)
\label{W eqn}
\end{eqnarray}

and

\begin{eqnarray}
\frac{dV(r)}{dr}=2\frac{d\Phi(r)}{dr}V(r) - e^{\Lambda(r)}\frac{W(r)}{r^2}
\label{V eqn}
\end{eqnarray}

For the fundamental ($f$) mode, $l$=2. The Eqs. \ref{W eqn} and \ref{V eqn} need to be solved by imposing certain boundary conditions at the center ($r$=0) and the surface ($r=R$) of the star. At $r$=0 the functions $W(r)$ and $V(r)$ behave as

\begin{eqnarray}
W(r)=Cr^{l+1}~~~~{\rm{and}}~~~V(r)=-Cr^l/l
\label{bc_center}
\end{eqnarray}

and at $r=R$ with vanishing perturbation the boundary condition of $W(r)$ and $V(r)$ is in the form of

\begin{eqnarray}
\omega^2r^2e^{\Lambda(R)-2\Phi(R)}V(R) + \frac{d\Phi(r)}{dr}\Bigg\rvert_{r=R}W(R) = 0
\label{bc_surface}
\end{eqnarray}

The coupled differential equations, Eqs.~(\ref{W eqn}) and (\ref{V eqn}), are integrated from the center to the surface of the star by considering an initial value of $\omega^2$. The value of $\omega^2$ is improved after each integration with the help of the Ridder's method until Eq.~(\ref{bc_surface}) is satisfied.

The dimensionless tidal deformability ($\Lambda$) is obtained in terms of the mass, radius and the tidal love number ($k_2$) following \cite{Hinderer:2007mb,Hinderer:2009ca}.


\section{Results}
\label{Results}

In \cite{Guha:2024pnn} we obtained the suitable range of $m_{\chi}$, corresponding to the range of $k_F^{\chi}$=(0.01 - 0.07) GeV, that can yield reasonable DMANSs configurations in terms of the various astrophysical constraints on the structural properties of NSs. For the four hadronic models considered in this work, the range of 
$m_{\chi}$ is as follows:

\begin{center}
NL3 : $m_{\chi}$=(1 - 30) GeV\\
GM1 : $m_{\chi}$=(0.3 - 20) GeV\\
DD2 : $m_{\chi}$=(0.5 - 15) GeV\\
DD-ME2 : $m_{\chi}$=(0.5 - 15) GeV
\end{center}

We first show the structural properties like mass, radius, tidal deformability and then the $f$-mode frequency of the DMANSs considering $k_F^{\chi}$=0.07 GeV for the range of $m_{\chi}$ corresponding to each hadronic model. 

\begin{figure}[!ht]
\centering
\subfloat[]{\includegraphics[width=0.49\textwidth]{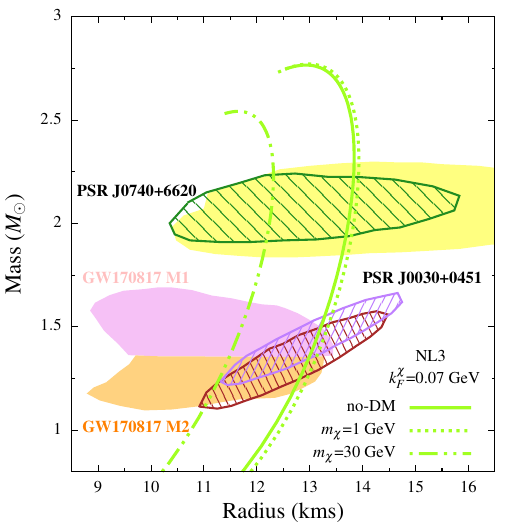}\protect\label{mr_NL3}}
\subfloat[]{\includegraphics[width=0.49\textwidth]{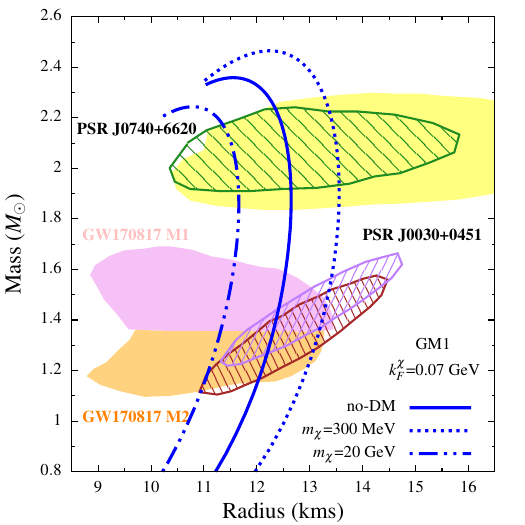}\protect\label{mr_GM1}}
\hfill
\subfloat[]{\includegraphics[width=0.49\textwidth]{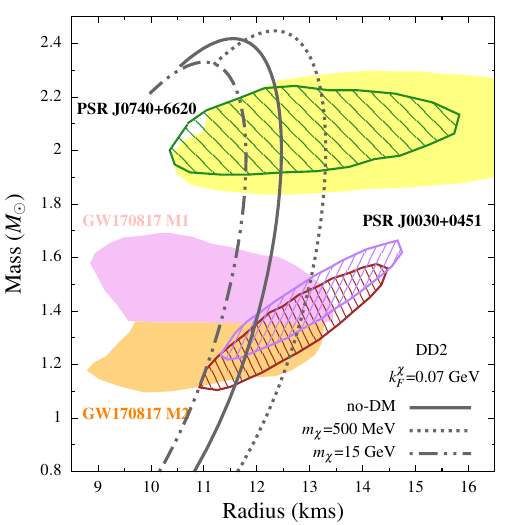}\protect\label{mr_DD2}}
\subfloat[]{\includegraphics[width=0.49\textwidth]
{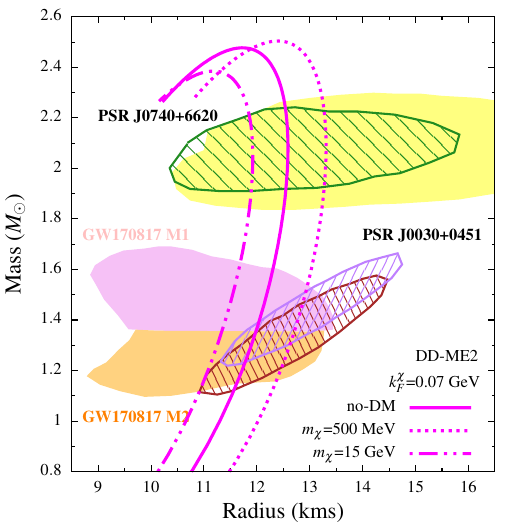}\protect\label{mr_DDME2}}
\caption{\it Variation of mass with radius of neutron stars with and without dark matter for $k_F^{\chi}$=0.07 GeV and different hadronic models. The observational limits imposed from the most massive pulsar PSR J0740+6620 ($M = 2.08 \pm 0.07 M_{\odot}$) \cite{Fonseca:2021wxt} and $R = 13.7^{+2.6}_{-1.5}$ km \cite{Miller:2021qha} or $R = 12.39^{+1.30}_{-0.98}$ km \cite{Riley:2021pdl} are also indicated. The constraints on $M-R$ plane prescribed from GW170817 \cite{LIGOScientific:2018cki} and the NICER experiment for PSR J0030+0451  \cite{Riley:2019yda, Miller:2019cac} are also compared.}
\label{mr}
\end{figure} 

\begin{figure}[!ht]
\centering
\subfloat[]{\includegraphics[width=0.49\textwidth]{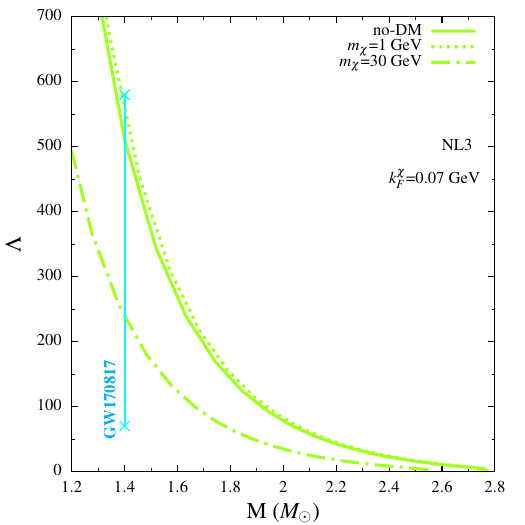}\protect\label{LamM_NL3}}
\subfloat[]{\includegraphics[width=0.49\textwidth]{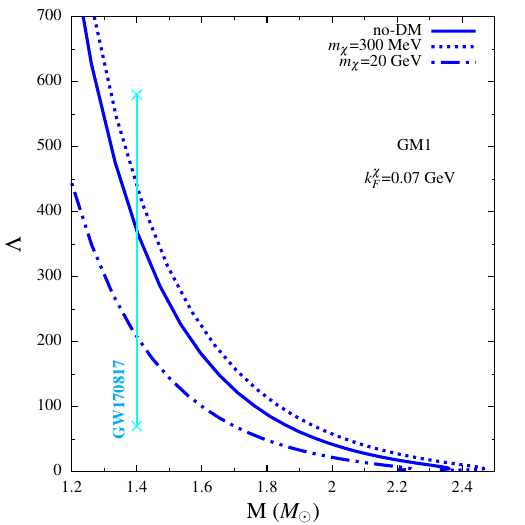}\protect\label{LamM_GM1}}
\hfill
\subfloat[]{\includegraphics[width=0.49\textwidth]{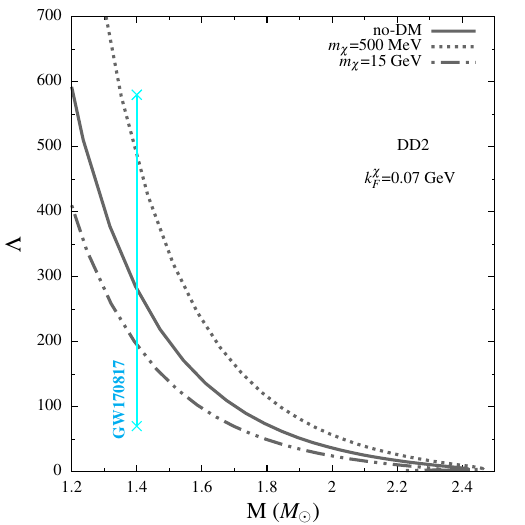}\protect\label{LamM_DD2}}
\subfloat[]{\includegraphics[width=0.49\textwidth]
{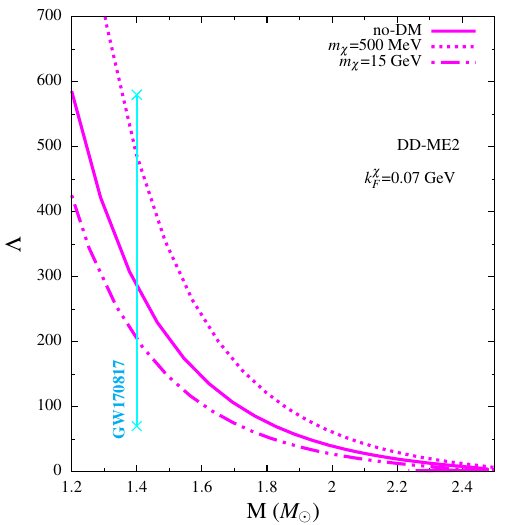}\protect\label{LamM_DDME2}}
\caption{\it Variation of tidal deformability with mass of neutron stars with and without dark matter  for $k_F^{\chi}$=0.07 GeV and different hadronic models. The constraint on $\Lambda_{1.4}$ from GW170817 \cite{LIGOScientific:2018cki} is also shown.}
\label{LamM}
\end{figure} 

In Figs. (\ref{mr}) and (\ref{LamM}) we display the variation of mass with respect to radius and the relationship of the tidal deformability with mass of DMANSs. We have also compared the case of pure hadronic stars i.e, the no-DM scenario. The presence of massive DM lowers both the  maximum mass and radius of the star. For $k_F^{\chi}$=0.07 GeV and for each hadronic model, the maximum and minimum values of $m_{\chi}$ are obtained for which all the constraints on the $M-R$ and $\Lambda-M$ planes are satisfied. For example, for the GM1 model, values of $m_{\chi}$ below 300 MeV do not satisfy the GW170817 data while $m_{\chi}>$ 20 GeV fail to satisfy NICER experiment for PSR J0030+0451. 

\begin{figure}[!ht]
\centering
\subfloat[]{\includegraphics[width=0.49\textwidth]{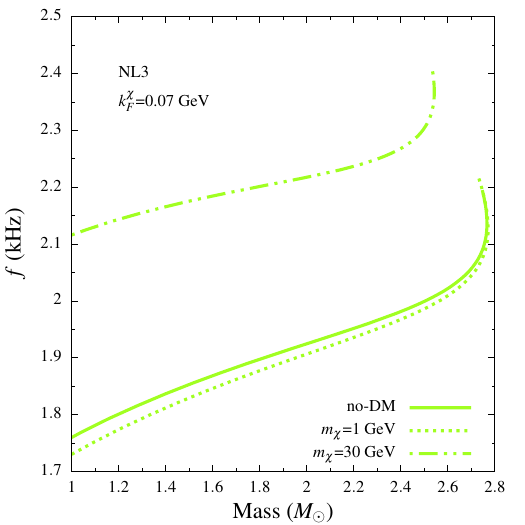}\protect\label{mf_NL3}}
\subfloat[]{\includegraphics[width=0.49\textwidth]{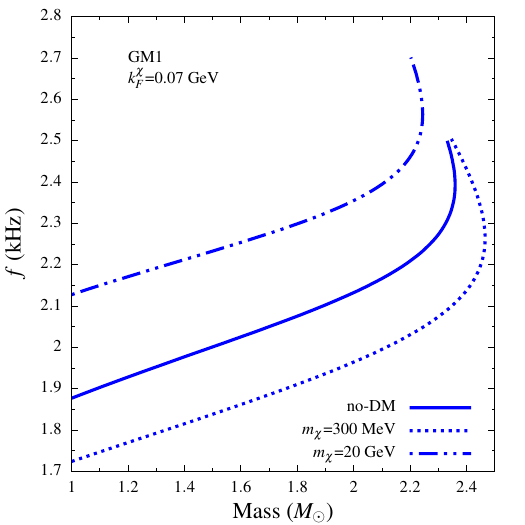}\protect\label{mf_GM1}}
\hfill
\subfloat[]{\includegraphics[width=0.49\textwidth]{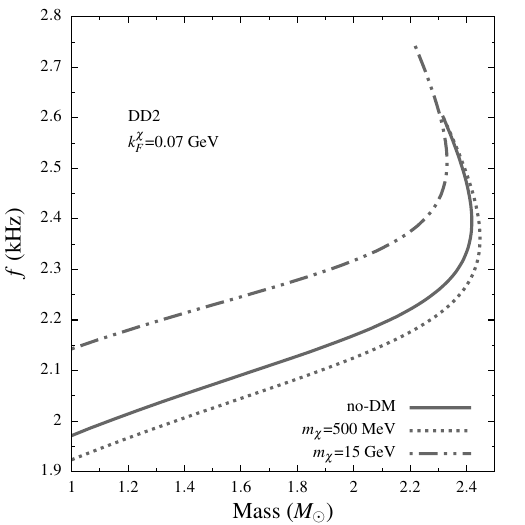}\protect\label{mf_DD2}}
\subfloat[]{\includegraphics[width=0.49\textwidth]
{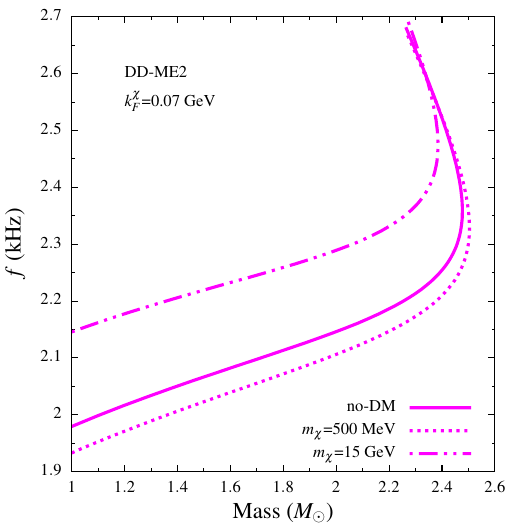}\protect\label{mf_DDME2}}
\caption{\it Variation of $f$-mode frequency with mass of neutron stars with and without dark matter for $k_F^{\chi}$=0.07 GeV and different hadronic models.}
\label{mf}
\end{figure} 

We next present in Fig. (\ref{mf}) the dependence of the $f$-mode frequency on mass of NSs without DM and with DM considering the fixed values of $k_F^{\chi}$ and $m_{\chi}$ respective to each hadronic model. The $f$-mode frequency increases with the mass of the star. Beyond the maximum mass $M_{max}$ of the star, the value of $f$ keeps increasing further. However, beyond $M_{max}$, the $M-R$ variation is unstable. Therefore, the values of $f$ in this region are not important and we consider $f_{max}$ as the value of $f$ corresponding to $M_{max}$. In the `no-DM scenario, the obtained values of $f_{max}$ in the present work are consistent with that obtained in \cite{Flores:2013yqa} for the NL3 and GM1 models and \cite{Ranea-Sandoval:2018bgu} for the NL3, GM1 and DD2 models. For instance, in case of the GM1 model, we find from Fig. (\ref{mf_GM1}) that in the `no-DM' scenario $f_{max}$=2.39 kHz corresponding to $M_{max}$=2.36 $M_{\odot}$ while in presence of DM, $f_{max}$=2.26 and 2.56 kHz corresponding to $M_{max}$=2.46 and 2.24 $M_{\odot}$ for $m_{\chi}$=300 MeV and 20 GeV, respectively. This also implies that the presence of DM has pronounced effects on the $f$ whose value is higher in case of massive DM. The value of $f$ depends on both the mass and radius of the DMANSs which are in turn dictated by $m_{\chi}$ via the DM admixed EoS. $f$ not only depends on the mass of the star but also on its radius. For example, in case of the NL3 model, we find from Fig. (\ref{mr_NL3}) that there is negligible difference in maximum mass of the star between the `no-DM' scenario and the case when $m_{\chi}$=30 GeV. The difference in radius between the two scenarios is also quite small, especially in the high mass region. The same effect is reflected in the behavior of $f$ in Fig. (\ref{mf_NL3}) where value of $f$, corresponding to the maximum mass of the star, is almost equal for the `no-DM' case and for $m_{\chi}$=30 GeV. There is, however, noticeable difference in the value of $f$ for the two scenarios at low mass regime which is a reflection of the behavior of radius in this regime.

Now for any particular hadronic model, the upper and the lower limits of $m_{\chi}$ are fixed for $k_F^{\chi}$=(0.01 - 0.07) GeV \cite{Guha:2024pnn}. This implies that in the presence of DM with a fixed $k_F^{\chi}$, the range of maximum mass configurations of the DMANSs is also fixed. The lower values of $m_{\chi}$ produces more massive DMANSs with respect to the radius as seen from Fig. (\ref{mr}). Consequently, the maximum mass of the DMANS for a given model becomes higher for the lower values of $m_{\chi}$. Therefore, $m_{\chi}^{max}$ for any model produces DMANS with the lowest value of the maximum mass for that model in the range $k_F^{\chi}$=(0.01 - 0.07) GeV. Let us denote this lowest value of the maximum mass as $M_{ll}^{DMANS}$. Similarly, $m_{\chi}^{min}$ yields DMANS with the highest value of the maximum mass for that model. We denote this highest value of the maximum mass as $M_{ul}^{DMANS}$. For example, for the GM1 model $m_{\chi}^{max}$=20 GeV gives $M_{ll}^{DMANS}$=2.24 $M_{\odot}$ while $M_{ul}^{DMANS}$=2.46 for $m_{\chi}^{min}$=300 MeV. Consequently, in the presence of DM, the upper and the lower limits of $f_{max}$ ($f$ corresponding to $M_{max}$), are also fixed for any particular model. Now, contrary to the values of the maximum mass of DMANSs, it is seen from Fig. (\ref{mf}) that massive DM gives rise to higher values of $f_{max}$ i.e., $m_{\chi}^{max}$ for any model gives the upper limit of the $f_{max}$ while $m_{\chi}^{min}$ produces the lower limit of $f_{max}$ for a fixed model. We denote the upper limit of $f_{max}$ as $f_{ul}^{DMANS}$ produced by $m_{\chi}^{max}$ and the lower limit of $f_{max}$ as $f_{ll}^{DMANS}$ given by $m_{\chi}^{min}$. As an example for the GM1 model, $m_{\chi}^{max}$=20 GeV gives  $f_{ul}^{DMANS}$=2.56 kHz while for $m_{\chi}^{min}$=300 MeV, $f_{ll}^{DMANS}$=2.26 kHz. Therefore, in the following table we show the values of $f_{ul}^{DMANS}$ and $f_{ll}^{DMANS}$ corresponding to $M_{ll}^{DMANS}$ and $M_{ul}^{DMANS}$ with $m_{\chi}^{max}$ and $m_{\chi}^{min}$, respectively.
 
\begin{table}[!ht]
\caption{The values of $f_{ll}^{DMANS}$ and $f_{ul}^{DMANS}$ for dark matter admixed neutron stars with corresponding $M_{ll}^{DMANS}$ and $M_{ul}^{DMANS}$ and $m_{\chi}^{min}$ and $m_{\chi}^{max}$ for $k_F^{\chi}$=0.07 GeV.}
\setlength{\tabcolsep}{10.0pt}
\begin{tabular}{ccccccc}
\hline
\hline
Model & $m_{\chi}^{min}$ & $M_{ul}^{DMANS}$ & $f_{ll}^{DMANS}$ & $m_{\chi}^{max}$ & $M_{ll}^{DMANS}$ & $f_{ul}^{DMANS}$\\
 & (GeV) & ($M_{\odot}$) & (kHz) & (GeV) & ($M_{\odot}$) & (kHz)\\
\hline
NL3     & 1   & 2.77 & 2.13 & 30 & 2.54 & 2.36\\
GM1     & 0.3 & 2.46 & 2.26 & 20 & 2.24 & 2.56\\
DD2     & 0.5 & 2.45 & 2.36 & 15 & 2.33 & 2.52 \\
DD-ME2  & 0.5 & 2.50 & 2.33 & 15 & 2.38 & 2.47 \\
\hline
\hline
\end{tabular}
\label{tab:max}
\end{table}

\begin{figure}[!ht]
\centering
\subfloat[]{\includegraphics[width=0.5\textwidth]{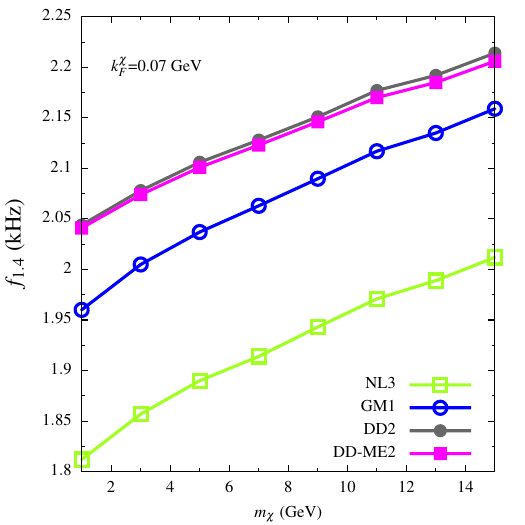}\protect\label{mchi_f1p4_0p06}}
\subfloat[]{\includegraphics[width=0.5\textwidth]{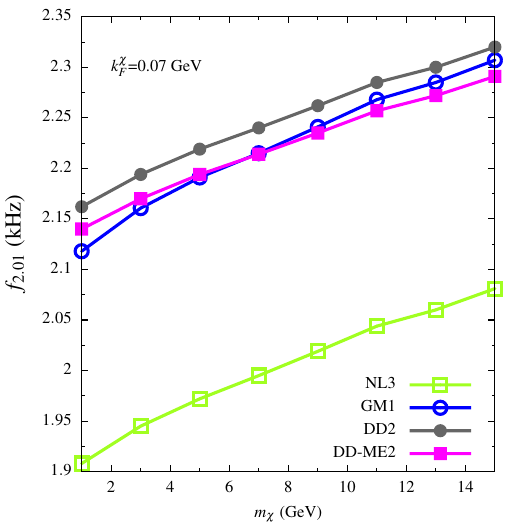}\protect\label{mchi_f2p01_0p06}}
\caption{\it Variation of $f_{1.4}$ and $f_{2.01}$  with mass of fermionic dark matter for $k_F^{\chi}$=0.07 GeV compared for different hadronic models.}
\label{mchi_f1p4_f2p01_0p06}
\end{figure}

Considering all the four hadronic models, we find that the common allowed range of $m_{\chi}$ is (1 - 15) GeV. Therefore, within this $m_{\chi}$ and fixed $k_F^{\chi}$=0.07 GeV, we next compare in Fig. (\ref{mchi_f1p4_f2p01_0p06}) the values of $f_{1.4}$ and $f_{2.01}$ for the DMANSs obtained with the different hadronic models. $f_{1.4}$ and $f_{2.01}$ are the values of $f$ corresponding to the DMANSs of mass 1.4 $M_{\odot}$ and 2.01 $M_{\odot}$, respectively. These values are important because one of the binary components associated with GW170817 has mass 1.4$M_{\odot}$ while the lower limit of mass of the most massive pulsar PSR J0740+6620, detected till date, is 2.01 $M_{\odot}$. The underlying RMF EoS plays a role in deciding the magnitude of $f$-mode. For example, for any particular value of $m_{\chi}$ and $k_F^{\chi}$, Fig. \ref{mchi_f1p4_0p06} shows that the value of $f_{1.4}$ in terms of the model, follows the sequence NL3$<$GM1$<$DD-ME2$<$DD2. Since a 1.4$M_{\odot}$ is obtained at low values of density, we find that at lower density the stiffness of the EoS in the ‘no-DM’ scenario follows the reverse trend as NL3$>$GM1$>$DD-ME2$>$DD2. Thus it can be said that stiffer EoS results in $f_{1.4}$ of lower magnitude. As expected, both $f_{1.4}$ and $f_{2.01}$ increase with $m_{\chi}$ for all the models. Figs. (\ref{mchi_f1p4_0p06}) (\ref{mchi_f2p01_0p06}) show that for any value of $m_{\chi}$, the values of both $f_{1.4}$ and $f_{2.01}$ are minimum for NL3 and maximum for DD2. This is because the values of $R_{1.4}$ and $R_{2.01}$ are maximum for NL3 and minimum for DD2. The values of both $f_{1.4}$ and $f_{2.01}$ are almost same for DD2 and DD-ME2 models for any value of $m_{\chi}$.

\begin{figure}[!ht]
\centering
\subfloat[]{\includegraphics[width=0.5\textwidth]{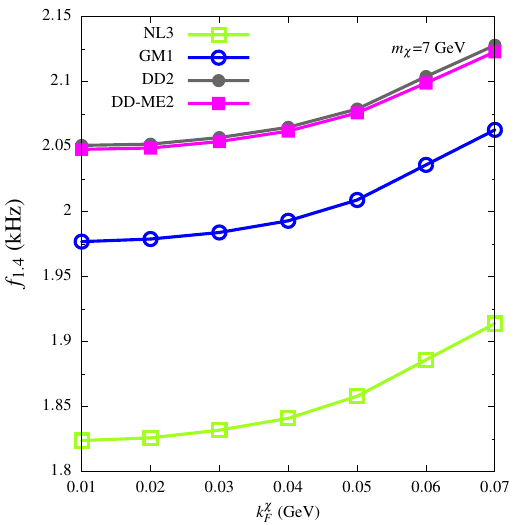}\protect\label{kfchi_f1p4_mchi7G}}
\subfloat[]{\includegraphics[width=0.5\textwidth]{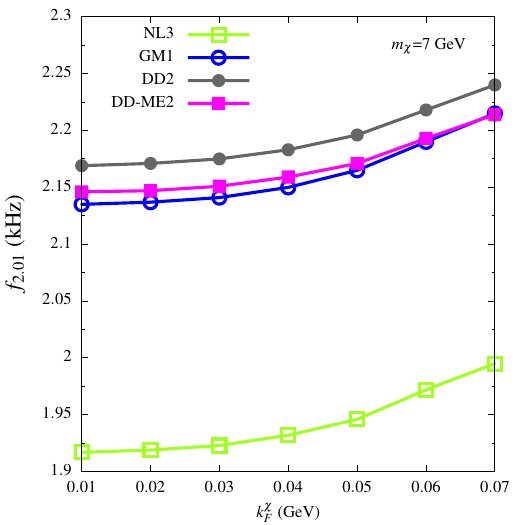}\protect\label{kfchi_f2p01_mchi7G}}
\caption{\it Variation of $f_{1.4}$ and $f_{2.01}$  with constant Fermi momentum of dark matter for $m_{\chi}$=7 GeV compared for different hadronic models.}
\label{kfchi_f1p4_f2p01_mchi7G}
\end{figure}

The common range of $m_{\chi}$=(1 - 15) GeV is not only valid for $k_F^{\chi}$=0.07 GeV but within a range of $k_F^{\chi}$=(0.01 - 0.07) GeV, as we obtained in \cite{Guha:2024pnn}. Therefore in Fig. (\ref{kfchi_f1p4_f2p01_mchi7G}) we now study the variation of $f_{1.4}$ and $f_{2.01}$ with respect to $k_F^{\chi}$ in the range of (0.01 - 0.07) GeV for a fixed value of $m_{\chi}$=7 GeV that falls within the acceptable common range of $m_{\chi}$=(1 - 15) GeV. The dependence of $f_{1.4}$ and $f_{2.01}$ on $k_F^{\chi}$ is compared for all the four hadronic models. The values of $f_{1.4}$ and $f_{2.01}$ increase with $k_F^{\chi}$ and similar to Fig. (\ref{mchi_f1p4_f2p01_0p06}), the values of $f_{1.4}$ and $f_{2.01}$ are maximum for DD2 model and minimum for NL3 for any particular value of $m_{\chi}$. Interestingly, from Figs. (\ref{mchi_f1p4_f2p01_0p06}) and (\ref{kfchi_f1p4_f2p01_mchi7G}) we observe that the variation of $f_{1.4}$ with both $m_{\chi}$ (Fig. \ref{mchi_f1p4_0p06}) and $k_F^{\chi}$ (Fig. \ref{kfchi_f1p4_mchi7G}) are obtained parallel for all the four hadronic models while in case of variation of $f_{2.01}$ the curves for the GM1 and DD-ME2 models intersect and overlap.

\begin{figure}[!ht]
\centering
\subfloat[]{\includegraphics[width=0.5\textwidth]{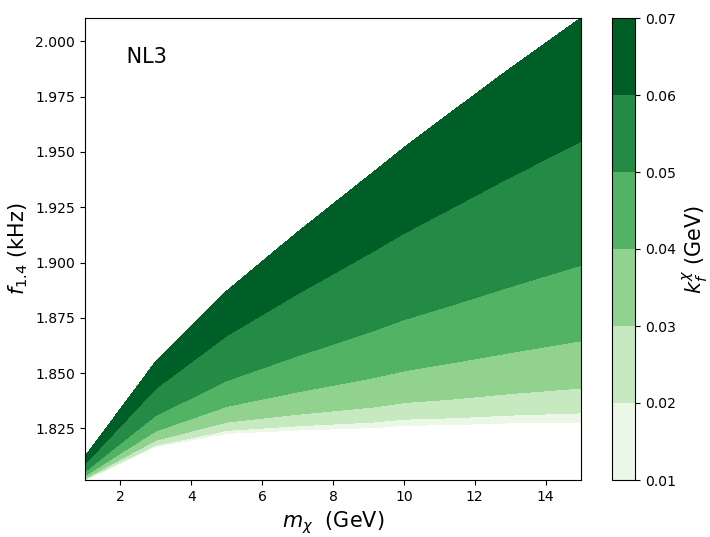}\protect\label{frequency_1p4_NL3}}
\hfill
\subfloat[]{\includegraphics[width=0.5\textwidth]{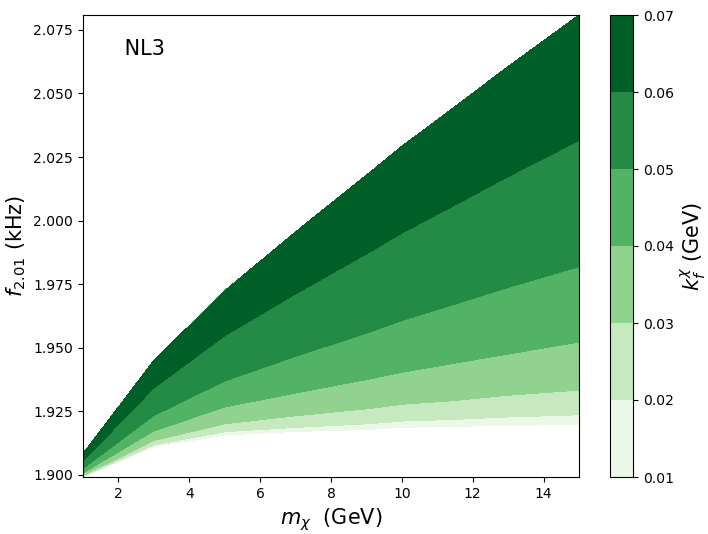}\protect\label{frequency_2p01_NL3}}
\hfill
\subfloat[]{\includegraphics[width=0.50\textwidth]{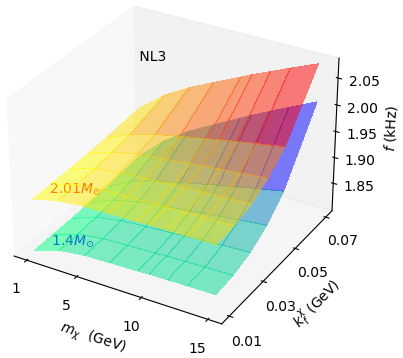}\protect\label{frequency_3D_NL3}}
\caption{\it Variation of $f_{1.4}$ and $f_{2.01}$  with $m_{\chi}$ and $k_F^{\chi}$ with NL3 hadronic model.}
\label{frequency_NL3}
\end{figure} 

\begin{figure}[!ht]
\centering
\subfloat[]{\includegraphics[width=0.5\textwidth]{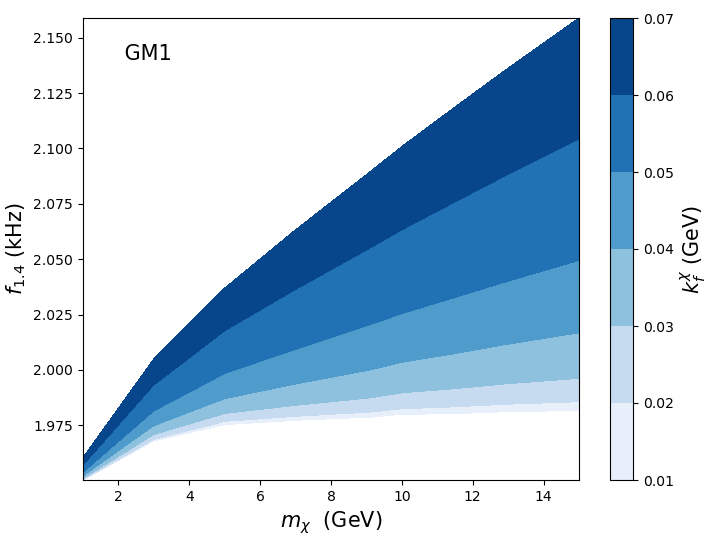}\protect\label{frequency_1p4_GM1}}
\hfill
\subfloat[]{\includegraphics[width=0.5\textwidth]{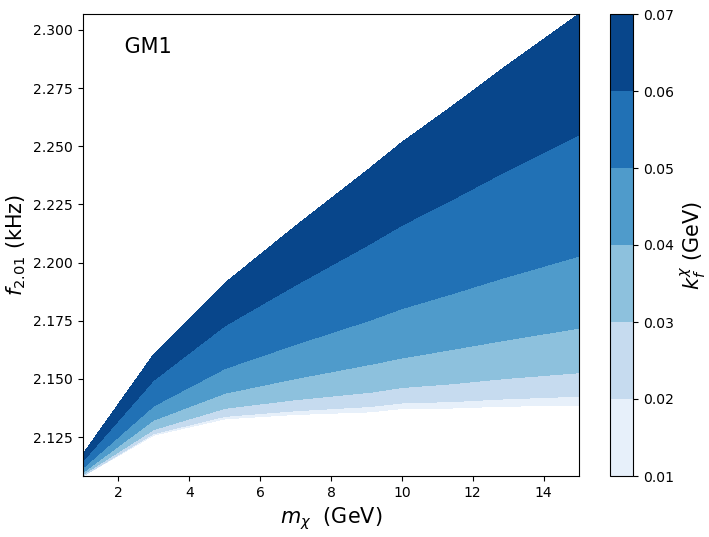}\protect\label{frequency_2p01_GM1}}
\hfill
\subfloat[]{\includegraphics[width=0.50\textwidth]{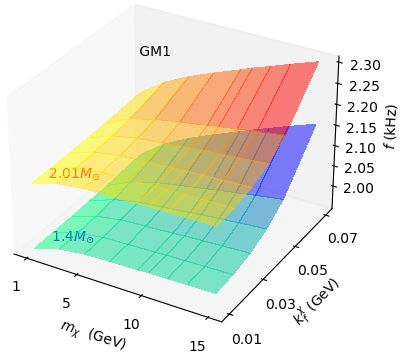}\protect\label{frequency_3D_GM1}}
\caption{\it Variation of $f_{1.4}$ and $f_{2.01}$  with $m_{\chi}$ and $k_F^{\chi}$ with GM1 hadronic model.}
\label{frequency_GM1}
\end{figure} 

\begin{figure}[!ht]
\centering
\subfloat[]{\includegraphics[width=0.5\textwidth]{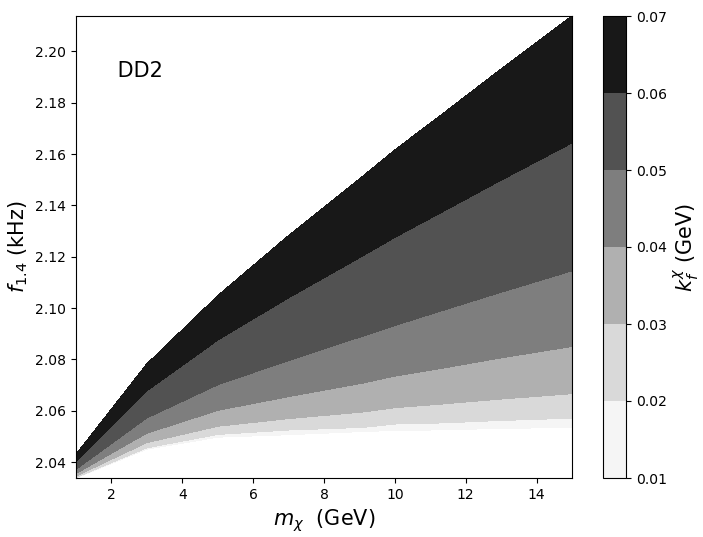}\protect\label{frequency_1p4_DD2}}
\hfill
\subfloat[]{\includegraphics[width=0.5\textwidth]{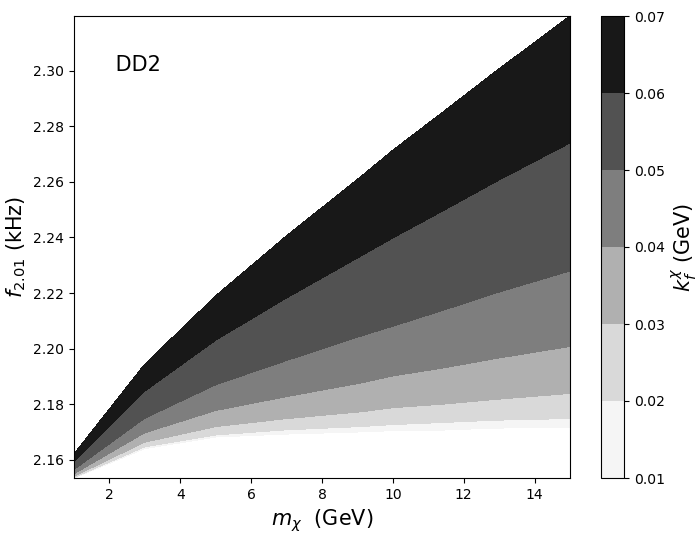}\protect\label{frequency_2p01_DD2}}
\hfill
\subfloat[]{\includegraphics[width=0.50\textwidth]{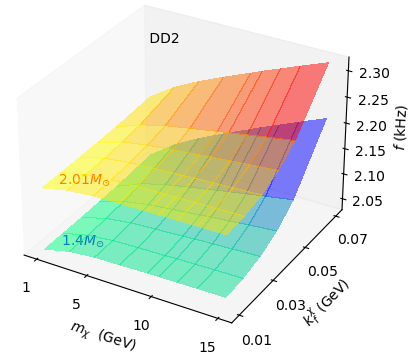}\protect\label{frequency_3D_DD2}}
\caption{\it Variation of $f_{1.4}$ and $f_{2.01}$  with $m_{\chi}$ and $k_F^{\chi}$ with DD2 hadronic model.}
\label{frequency_DD2}
\end{figure} 

\begin{figure}[!ht]
\centering
\subfloat[]{\includegraphics[width=0.5\textwidth]{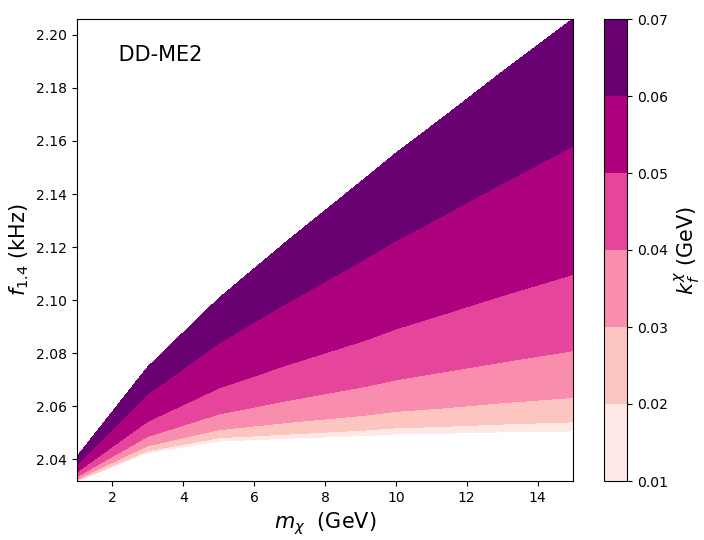}\protect\label{frequency_1p4_DDME2}}
\hfill
\subfloat[]{\includegraphics[width=0.5\textwidth]{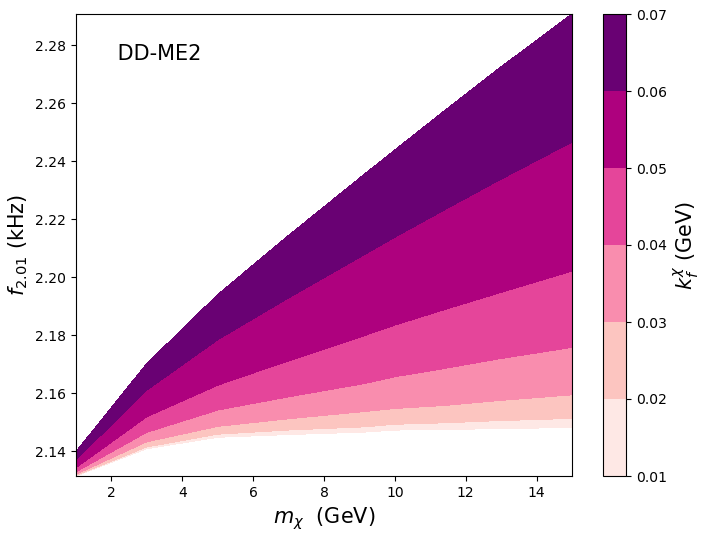}\protect\label{frequency_2p01_DDME2}}
\hfill
\subfloat[]{\includegraphics[width=0.50\textwidth]{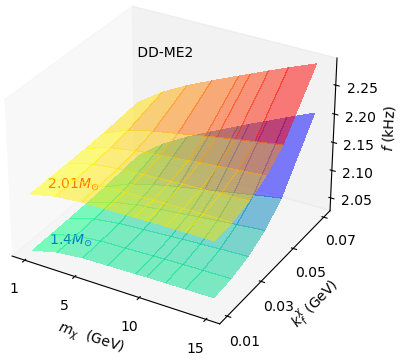}\protect\label{frequency_3D_DDME2}}
\caption{\it Variation of $f_{1.4}$ and $f_{2.01}$  with $m_{\chi}$ and $k_F^{\chi}$ with DD-ME2 hadronic model.}
\label{frequency_DDME2}
\end{figure} 

We finally try to understand the complete effect of the presence of DM on the values of $f_{1.4}$ and $f_{2.01}$ for the four individual hadronic models NL3, GM1, DD2 and DD-ME2 in Figs. (\ref{frequency_NL3}), (\ref{frequency_GM1}), (\ref{frequency_DD2}) and (\ref{frequency_DDME2}), respectively. The complete effect of DM can be seen from the combined effect of $m_{\chi}$ and $k_F^{\chi}$ which is studied in these figures. We span the common allowed range of $m_{\chi}$=(1 - 15) GeV fully and also the total range of $k_F^{\chi}$=(0.01 - 0.07) GeV in the following figures. Figs. (\ref{frequency_1p4_NL3}), (\ref{frequency_2p01_NL3}), (\ref{frequency_1p4_GM1}), (\ref{frequency_2p01_GM1}), (\ref{frequency_1p4_DD2}), (\ref{frequency_2p01_DD2}), and (\ref{frequency_1p4_DDME2}), (\ref{frequency_2p01_DDME2}) show that for any particular hadronic model $f_{1.4}$ and $f_{2.01}$ increase with $m_{\chi}$. This increasing behavior is further escalated with the larger values of $k_F^{\chi}$. 

Each band represent the variation of $f_{1.4}$ and/or $f_{2.01}$ with respect to $m_{\chi}$ for the value of $k_F^{\chi}$ within an interval of 0.01 GeV. The interface of any two consecutive bands represents $f_{1.4}$ and/or $f_{2.01}$ with respect to $m_{\chi}$ for a fixed value of  $k_F^{\chi}$ which is denoted in the color bar. It is interesting to note that for decreasing $k_F^{\chi}$, there is monotonic squeezing of the bands. This is because at very low $k_F^{\chi}$, the DM population in terms of DM mass density $\rho_{\chi}$ is too low, irrespective of $m_{\chi}$. Under such circumstances, the findings related to the structural properties of the star are very similar to that of the `no-DM' scenario, as also seen in \cite{Guha:2024pnn}. For example, for the DD-ME2 model, the values of $f_{1.4}$ and $f_{2.01}$ for $k_F^{\chi}$=0.01 GeV are approximately 2.05 and 2.15 kHz, respectively as seen from Figs. \ref{frequency_1p4_DDME2} and \ref{frequency_2p01_DDME2}. These values of $f_{1.4}$ and $f_{2.01}$ are very close to that obtained in the `no-DM' scenario for DD-ME2 model as seen from Fig. \ref{mf_DDME2}.

In Figs. (\ref{frequency_3D_NL3}), (\ref{frequency_3D_GM1}), (\ref{frequency_3D_DD2}), and (\ref{frequency_3D_DDME2}) we depict the 3-dimensional representation of the variation of $f_{1.4}$ and $f_{2.01}$ with respect to both $m_{\chi}$ and $k_F^{\chi}$. In such representation, we obtain the values of $f_{1.4}$ and $f_{2.01}$ as two separate non-overlapping planes. From Figs. (\ref{frequency_NL3}), (\ref{frequency_GM1}), (\ref{frequency_DD2}) and (\ref{frequency_DDME2}) we also provide the range of $f_{1.4}$ and $f_{2.01}$ of the DMANSs for the accepted range of $m_{\chi}$ corresponding to the range of $k_F^{\chi}$. For example, for the DD2 model, $f_{1.4}^{DMANS}$=(2.03 - 2.16) kHz and $f_{2.01}^{DMANS}$=(2.15 - 2.32) kHz.

\begin{figure}[!ht]
\centering
\subfloat[]{\includegraphics[width=0.5\textwidth]{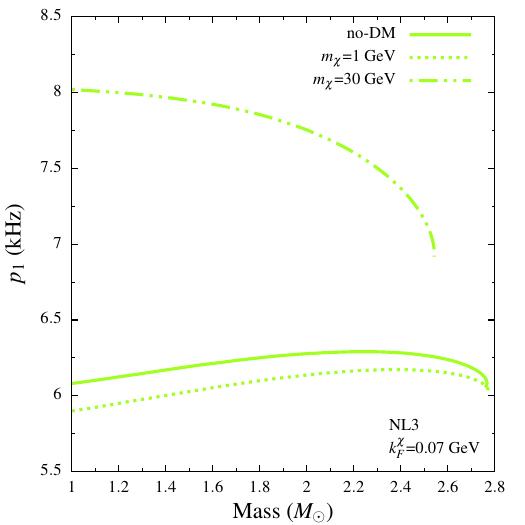}\protect\label{mp_mchi_NL3}}
\subfloat[]{\includegraphics[width=0.5\textwidth]{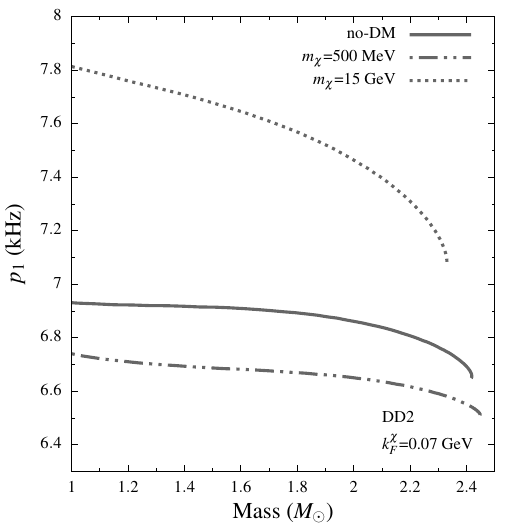}\protect\label{mp_mchi_DD2}}
\caption{\it Variation of $p_1$-mode frequency with mass of neutron stars with and without dark matter for $k_F^{\chi}$=0.07 GeV and different hadronic models.}
\label{mp_mchi}
\end{figure}

{We calculate the $p_1$-mode frequency of the NSs without DM and with DM considering the fixed value of $k_F^{\chi}$=0.07 GeV. For DMANSs we choose to work with the maximum and minimum values of $m_{\chi}$ respective to different hadronic models. In case of the $p$-mode, $l$=2, similar to the $f$-mode scenario. However, the number of nodes for the $p$-mode is $n$=1, unlike the $f$-mode which has no node. In Fig. \ref{mp_mchi} we show the variation of the $p_1$-mode frequency with respect to the mass of the of NSs with and without DM for the two hadronic models NL3 (Fig. \ref{mp_mchi_NL3}) and DD2 (Fig. \ref{mp_mchi_DD2}). We know that unlike the $f$-mode frequencies, which are not much affected by the crust \cite{VasquezFlores:2017tkp, Pradhan:2020amo, Zheng:2023oba, Thapa:2023grg}, the $p$-mode is quite sensitive to the low-density crustal EoS \cite{Kunjipurayil:2022zah}. As stated earlier that we have not considered the crustal effects of the NSs. Therefore, our results of the magnitude of the $p_1$-mode frequency of the NSs without DM are different (a bit larger) than that obtained in \cite{Flores:2013yqa, Ranea-Sandoval:2018bgu} especially at low mass (density). However, the value of ${p_1}_{max}$ (corresponding to $M_{max}$) do not differ much from that obtained in \cite{Flores:2013yqa, Ranea-Sandoval:2018bgu} as the deviation is noticed only at the second significant figure. In the present work we obtain the values of ${p_1}_{max}$ for the NSs without DM as 6.08 kHz for NL3 model and 6.66 kHz for the DD2 model while in \cite{Flores:2013yqa} ${p_1}_{max}$ is obtained as $\sim$6 kHz for the NL3 model and in \cite{Ranea-Sandoval:2018bgu} the value of ${p_1}_{max}$ is $\sim$6.5 kHz. It is observed from both the Figs. \ref{mp_mchi_NL3} and \ref{mp_mchi_DD2} that for the cases of `no-DM' and the DMANSs with minimum value of $m_{\chi}$, the $p_1-M$ variation initially follows a flat or slightly increasing trend that decreases after a certain value of mass (density). For the maximum value of $m_{\chi}$, the variation of $p_1$ shows an overall decreasing trend along the mass (density) profile of the star. In the present work we obtain the range of ${p^{DMANS}_{1max}}$ as (6.03 - 6.94) kHz for the NL3 model and (6.49 - 7.04) kHz for the DD2 model.}


\clearpage 

\section{Summary and Conclusion}
\label{Conclusion}

The feeble interaction between the fermionic DM ($\chi$) with the hadronic matter is introduced through a dark scalar ($\phi$) and a dark vector ($\xi$) boson as mediators in \cite{Guha:2024pnn} where $m_{\chi}$ is related to $m_{\phi}$ and $m_{\xi}$ via the constraint from Bullet cluster while the coupling strengths of $\phi$ and $\xi$ with $\chi$ are obtained using the bound from the present day relic abundance. In the same paper, we prescribed for different hadronic models, the allowed range of $m_{\chi}$ that can help the DMANS configurations to satisfy all the observational and astrophysical constraints on the structural properties like the mass, radius and tidal deformability of compact stars.

In the present work, we utilize the allowed range of $m_{\chi}$ to study the effects of such feeble interaction of DM with the hadronic star matter on the non-radial $f$-mode oscillation of the DMANSs considering the Cowling approximation. In terms of $m_{\chi}$ and $k_F^{\chi}$, we found that DM has profound influence in determining the $f$-mode frequency of NS oscillation. For a fixed value of $k_F^{\chi}$, we also determined the possible range of $f^{DMANS}$ corresponding to that of $m_{\chi}$ for four particular hadronic RMF models. For instance we obtain for $k_F^{\chi}$=0.07 GeV the range of $f^{DMANS}$ for the following hadronic models as follows:

\begin{center}
NL3 : $f_{max}^{DMANS}$=(2.13 - 2.36) kHz\\
GM1 : $f_{max}^{DMANS}$=(2.26 - 2.56) kHz\\
DD2 : $f_{max}^{DMANS}$=(2.36 - 2.52) kHz\\
DD-ME2 : $f_{max}^{DMANS}$=(2.33 - 2.47) kHz
\end{center}

So in the framework of our DMANSs in the present work, the range of $f_{max}^{DMANS}$=(2.13 - 2.56) kHz considering all the four hadronic models.

Since the properties of 1.4 and 2.01 $M_{\odot}$ are of current interest, we particularly studied both the individual and combined dependence of $f_{1.4}$ and $f_{2.01}$ on both $m_{\chi}$ and $k_F^{\chi}$. We found that the variation of $f_{1.4}$ and $f_{2.01}$ increase with $m_{\chi}$, which is further boosted by large values of $k_F^{\chi}$. For the four hadronic models, the range of $f_{1.4}$ and $f_{2.01}$ of DMANS with respect to the acceptable range of $m_{\chi}$ corresponding to the range of $k_F^{\chi}$ is tabulated below.
 
\begin{center}
NL3 : $f_{1.4}^{DMANS}$=(1.80 - 1.95) kHz\\
GM1 : $f_{1.4}^{DMANS}$=(1.95 - 2.16) kHz\\
DD2 : $f_{1.4}^{DMANS}$=(2.03 - 2.16) kHz\\
DD-ME2 : $f_{1.4}^{DMANS}$=(2.03 - 2.20) kHz
\end{center}

\begin{center}
NL3 : $f_{2.01}^{DMANS}$=(1.90 - 2.08) kHz\\
GM1 : $f_{2.01}^{DMANS}$=(2.11 - 2.31) kHz\\
DD2 : $f_{2.01}^{DMANS}$=(2.15 - 2.32) kHz\\
DD-ME2 : $f_{2.01}^{DMANS}$=(2.13 - 2.29) kHz
\end{center}

Thus in the present work, considering all the four hadronic models, $f_{1.4}^{DMANS}$ and $f_{2.01}^{DMANS}$ lie in the range (1.80 - 2.20) kHz and (1.90 - 2.32) kHz, respectively.

It should be mentioned that we have not included the effects of the NS crust in \cite{Guha:2024pnn} while obtaining the range of $m_{\chi}$ due to the absence of a suitable DM admixed crust EoS. Therefore, in future if a suitable DM admixed NS crust EoS can be obtained with a technique consistent with that used for obtaining the EoS of the core of the DMANSs, then it can be employed to re-calculate the gross properties of the DMANSs. In that case, the range of $m_{\chi}$ (consistent with the various astrophysical constraints) will be modified due to modification in the radii of the DMANSs. Consequently, the range of $f_{max}^{DMANS}$, $f_{1.4}^{DMANS}$, and $f_{2.01}^{DMANS}$ will be modified according to the modified range of $m_{\chi}$. 

 {We have also calculated the values of $p_1$-mode frequency of the NSs with and without DM. We found that in absence of the crustal effects, the values of $p_1$-mode in the `no-DM' case, are larger than that generally obtained in literature, especially at low density. Unlike the $p_1$-mode, the effects of the crust EoS is not remarkable on the $f$-mode of NSs. Therefore, in the present work we mainly focus on the variation and magnitude of the $f$-mode both in presence and absence of DM. Moreover, $f$-mode is of special interest because it is the most prominent oscillation frequency and thus compared to the higher order modes, it is more likely to be measured by the upcoming facilities like the Einstein Telescope, the Cosmic Explorer, and the LIGO O4 run.}
 
 Also, in the present work the non-radial oscillation frequencies like the $f$ and $p_1$ mode frequencies of the NSs with and without DM, are calculated in the Cowling approximation neglecting the metric perturbations in spacetime. Therefore, it will be interesting to calculate the oscillation frequencies in full GR conditions that include the spacetime oscillation as well. Such consideration will also give us the opportunity to calculate the $w$-mode frequency along with the $f$ and $p_1$ mode frequencies.


\section*{Acknowledgements}

Work of DS is supported by the NRF Research Grants (No. 2018R1A5A1025563). Work of AG is supported by the National Research Foundation of Korea, grant funded by Korea Government (MSIT) (RS-2024-00356960).


\bibliography{ref}

\end{document}